# Electromagnetic Compatibility of Power Converters

*A. Charoy*
AEMC, Sassenage, France

**Abstract**
This paper describes the main challenges related to power converters in a scientific environment. It proposes some methods of EMC analysis, design, measurement, and EMC troubleshooting.

## 1   Introduction

This report details and completes the Baden presentation on May 9, 2014.

Electromagnetic compatibility (EMC) is, broadly speaking, nothing but a control of electrical signals. Power converters, including uninterruptible power supplies (UPS), are particularly concerned with EMC.

Converters generate high-level conducted disturbances, mainly by large converters. A converter may be upset by its own disturbances (self-immunity) or disturb its environment (scientific equipment for instance), usually due to common-mode (CM) currents.

Converters (even low-power ones) may also radiate high-frequency electromagnetic energy that can upset nearby equipment (near-field coupling) or radio receivers (far-field coupling).

### 1.1   Beware of unreasonable EMC standards

A scientific environment looks like an industrial one, so the relevant standards use industrial EMC limits. The conducted emission limits of many EMC standards for large equipment (UPS, inverters, speed drives, arc welders, lifts…) are objectively too high, especially for equipment with rated current over 100 A.

**Mains terminal disturbance voltage limits for class A equipment measured on a test site**

| Frequency band | Class A equipment limits dB(µV) | | | | | |
|---|---|---|---|---|---|---|
| | Group 1 | | Group 2 | | Group 2 [a] | |
| MHz | Quasi-peak | Average | Quasi-peak | Average | Quasi-peak | Average |
| 0,15 – 0,50 | 79 | 66 | 100 | 90 | 130 | 120 |
| 0,50 – 5 | 73 | 60 | 86 | 76 | 125 | 115 |
| 5 – 30 | 73 | 60 | 90 / Decreasing linearly with logarithm of frequency to 70 | 80 / 60 | 115 | 105 |

[a]  Mains supply currents in excess of 100 A per phase when using the CISPR voltage probe or a suitable V-network (LISN or AMN).

**Fig. 1:** Example of a defective standard (here EN 62040-2 for UPS)

As can be seen in Fig. 1, the allowed relaxation in quasi-peak detection for a UPS over 100 A per phase is more than 50 dB higher (i.e., more than 350 times in amplitude) than for a UPS with a rated current up to 16 A. Moreover, the level of CM conducted emission up to 30 MHz corresponds to a value over 60 dB (1000 times) above the limit of radiated level over 30 MHz!

### 1.2 Conducted emission levels are frequently too high

A reasonable conducted emission limit for an industrial environment is CISPR class A. This limit is rather easy to meet up to 100 A per phase. For equipment with currents larger than 100 A per phase, class A + 10 dB is a reasonable objective (both in quasi-peak and average detection).

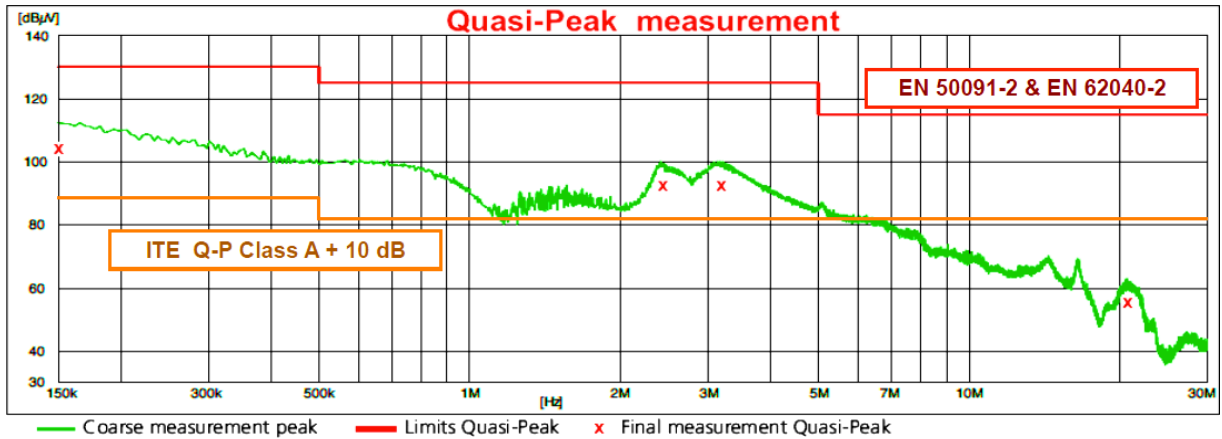

**Fig. 2:** Example of excessive conducted emission (80 kVA UPS, quasi-peak detection)

### 1.3 Beware of low-frequency band (from 2 kHz to 150 kHz)

Up to 2 kHz, the harmonic standard limits the emission levels to very low levels: no trouble.

Over 150 kHz, to meet a reasonable standard avoids most troubles.

But from 2 kHz to 150 kHz, there is still no limit for civilian standards. This band is jokingly called 'the Wild West'! Most of the troubles we meet on site appear at the switching frequency, between 4 kHz and 20 kHz. A CISPR draft is under consideration to complete the conducted voltage limit from 50 Hz to 150 kHz.

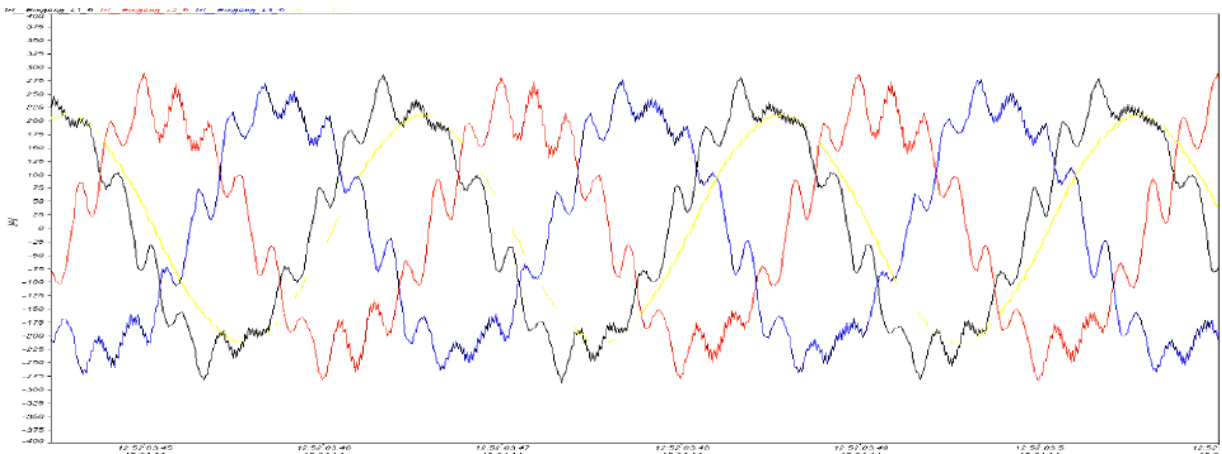

**Fig. 3:** Example of generated current (here the switching frequency is 5 kHz)

Pending normalized emission level, the ITER Organization at Cadarache specified a conducted emission limit based on a military standard (MIL-STD 461) for any cable (including signal cables).

This test, which uses a current probe, is easy to implement on site. The level fits CISPR limit current per phase over 150 kHz.

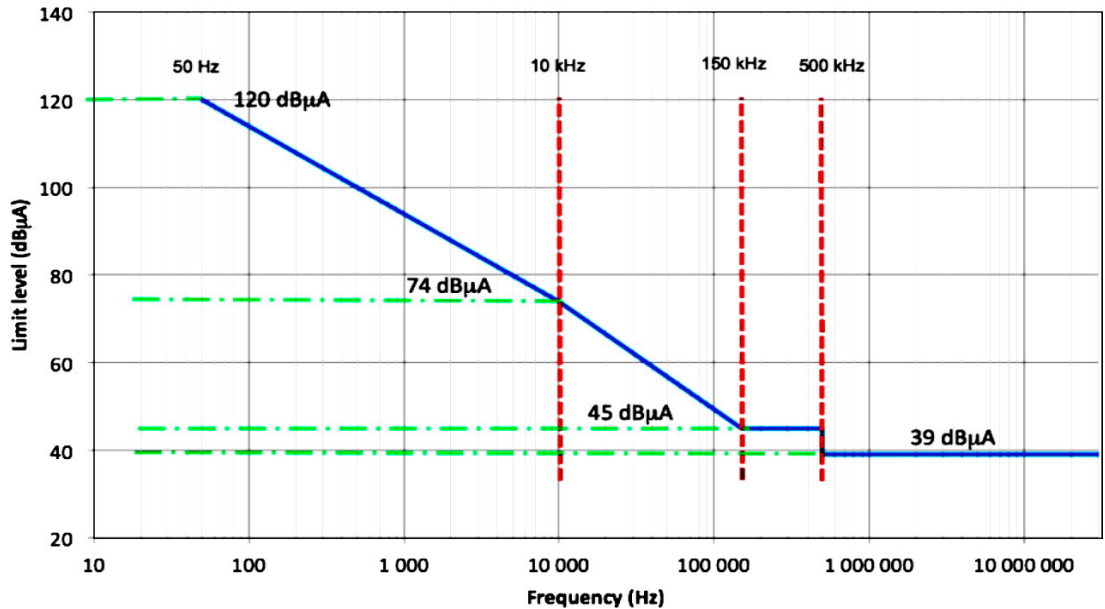

**Fig. 4:** Conducted emission limit for ITER Facility (here below a rated current of 1 A)

### 1.4 A missing immunity test

The generic immunity standard for an industrial environment, EN IEC 61000-6-2, specifies many (too many perhaps) immunity tests: electrostatic discharge, electromagnetic field immunity over 80 MHz, electrical fast transient in burst (an excellent revealing test), surges, CM immunity from 150 kHz to 80 MHz, voltage drops, and voltage fluctuations. But no immunity test is required in the frequency band from d.c. to 150 kHz. An abnormal vulnerability in this band may be ignored.

Yet a standard exists: IEC 61000-4-16. Cenelec plans to add this test for industrial environments.

## 2 EMC concerns of power converters

### 2.1 The negative impedance behaviour

Most d.c./d.c. converters have a high-voltage input dynamic range, frequently larger than a factor of 3. This means that an input-voltage reduction creates an input-current increase. This is called a 'negative impedance' phenomenon; it should not be confused with a possible transient output overvoltage.

The consequences of this—now well-known—behaviour, when the power supply distribution is not very low, can be varied:

- no start,
- start, but oscillates with the wrong output voltage (usually too low),
- output voltage seems correct, but the maximum output current is reduced,
- switching frequency instability, which can lead to destruction of the converter.

Fixes may be simple:

- add a larger input capacitor;

- reduce the power distribution impedance (using several pairs in parallel, for instance);
- reduce the converter regulation bandwidth.

## 2.2 Immunity of an optocoupler

Optocouplers and opto-isolators are components commonly used in power converters. Incredible as it may seem for electrical isolation components, the immunity of analogue optocouplers to CM disturbances is never specified (but in d.c.). Most digital optocouplers are specified for maximum d$V$/d$t$. Let us read the data sheet and comply with the limits, especially the maximum absolute ratings.

Many MOSFETs or IGBTs have a switching slope from 20 kV/$\mu$s to 50 kV/s, while most isolators are only specified up to a voltage rate immunity of 10–15 kV/$\mu$s. This doesn't lead to a destruction, but creates high-frequency oscillations that increase switching loss and radiated emissions, and may compromise reliability (due to IGBT second breakdown voltage, for instance).

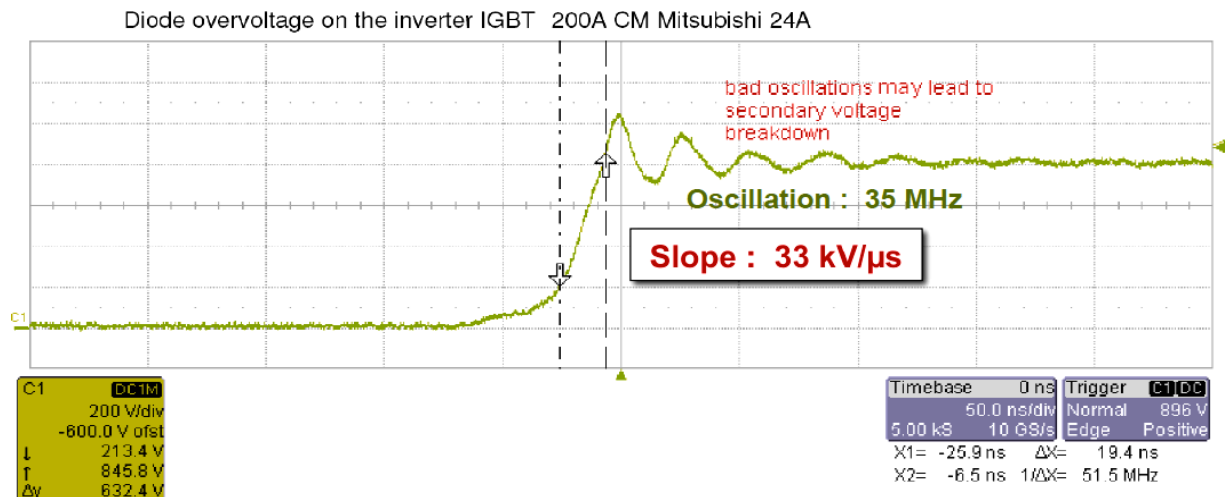

**Fig. 5:** Oscillation due to a voltage slope higher than opto-driver immunity (typically 15 kV/m)

## 2.3 Immunity and on-site fixes

The immunity of a power converter is usually correct. Nevertheless, some concern may occur regarding:

- cross-conduction of both transistors of the same arm of the H-bridge,
- instabilities or regulation faults (usually due to CM current on internal cables),
- transmission failure (usually due to CM current on transmission cable).

Fortunately, effective fixes—without drawback risks—can be applied on site:

- adding bonds between chassis or cabinet ground,
- using a shielded cable with its braid grounded at both ends,
- connecting the unused wires of a cable to chassis ground at both ends,
- adding a ferrite toroid (of a few turns) over a sensitive cable (or flat ribbon cable).

It is easy and revealing to measure the CM current on cables. For large converters, the usual useful bandwidth of the current clamp is 5–30 MHz. The maximum peak-to-peak current is typically 2 A, but—for a pretty good margin—we recommend limiting this value at 400 mA.

## 2.4 Input-to-output common-mode current for an isolated power supply

Most d.c./d.c. converters are isolated, using an insulating transformer. Despite this galvanic insulation, some common-mode current can flow through the stray capacitance of the transformer.

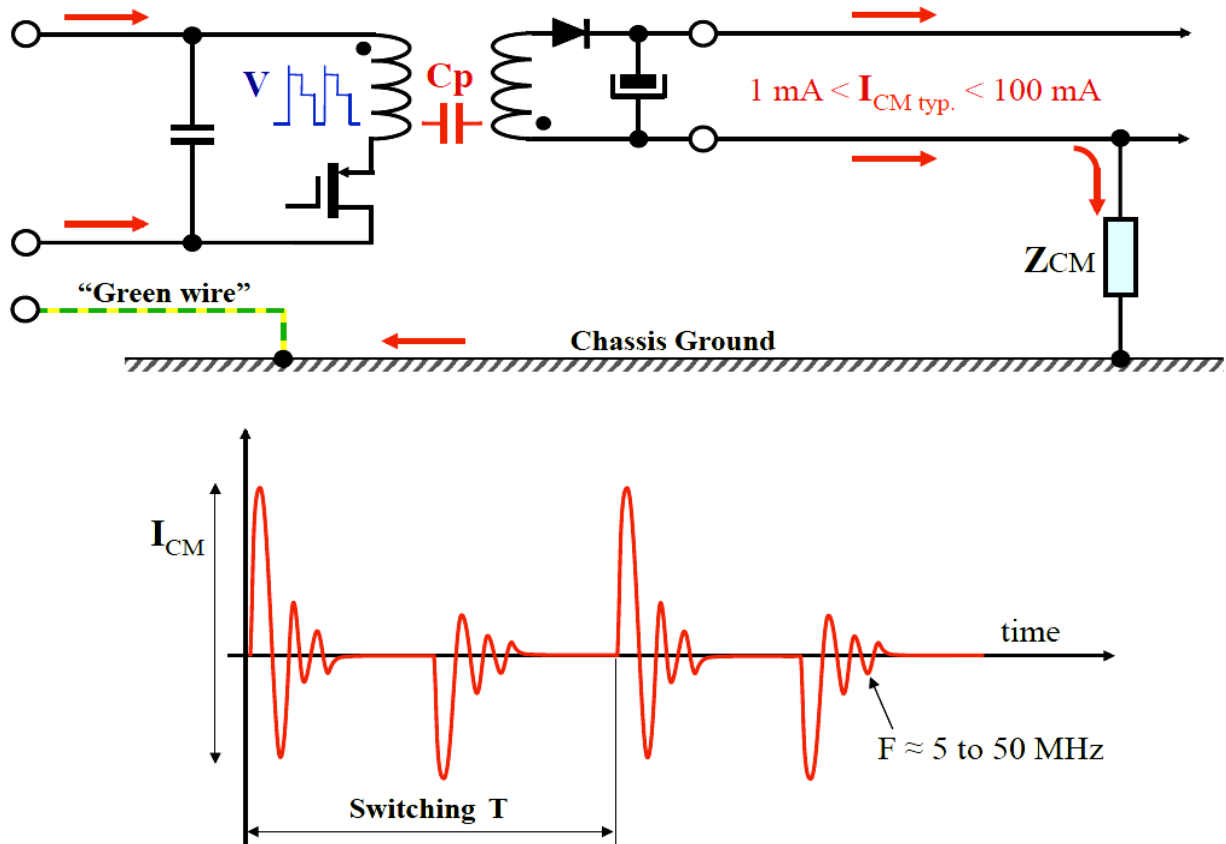

**Fig. 6:** Principle of primary-to-secondary CM current (typical waveform)

The amplitude of the CM current generated by an isolated d.c./d.c. converter may be as low as 1 mA (for a properly filtered power supply), or as large as 100 mA (for a poorly filtered d.c./d.c. converter). The waveform is a damped sine wave with a pseudo-frequency of 5–10 MHz for a large Switched Mode Power Supply (SMPS), reaching 25–50 MHz for a small SMPS. The repetition frequency is twice the fundamental frequency of the SMPS.

This CM current may disturb any sensitive equipment in the vicinity, starting with the supplied circuit itself. A simple solution to reduce this current is to add a CM choke or, if possible, to add capacitors from the output to the chassis ground. The total length of this connection—including the length of the metal column—must remain as short as practicable (less than 2 cm if possible).

## 2.5 Input-to-output CM current of an H-bridge

A single or three-phase H-bridge create large amount of common noise voltage. The CM impedance of an H-bridge is very low, so the external ground loop impedance limits the CM current.

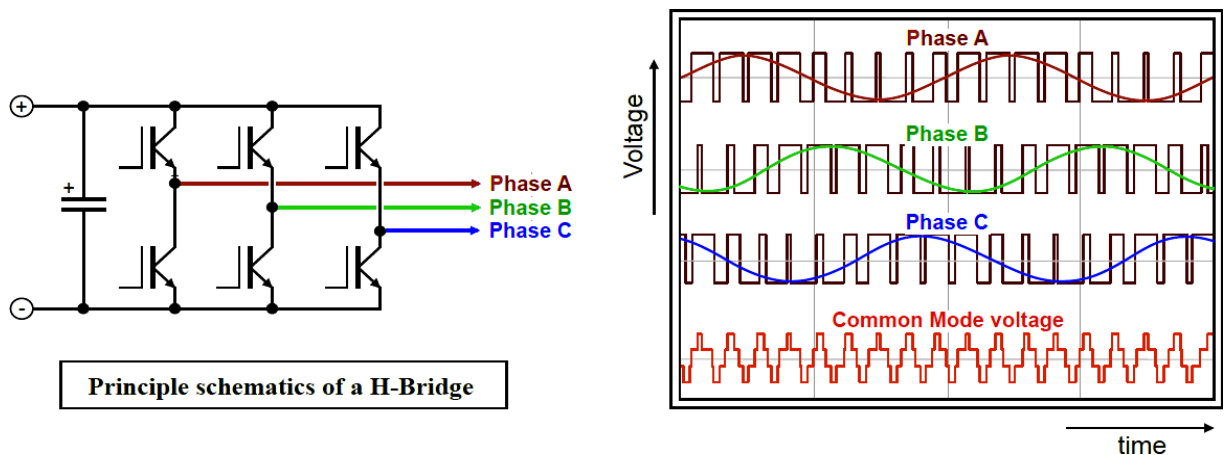

**Fig. 7:** Principle of a CM voltage generated by an H-bridge (here a three-phase H-bridge)

The amplitude of the CM current generated by the H-bridge may be as low as few milliamps (for a properly shielded output cable), or as large as 10 A (for a high-power motor drive). The waveform is a damped sine wave with a pseudo-frequency of 2–5 MHz for a large H-bridge, reaching 10–25 MHz for a small one.

The two common solutions to limit CM current are:

- to shield the output cable (connected to the chassis ground at both ends),
- to filter the output (but then a large filter is needed).

### 2.6 Common-mode disturbances depending on the topology

Common-mode disturbances emitted by a power converter depend greatly on its topology. Three main architectures can be analysed.

#### 2.6.1 Case 1: The output of the converter is grounded

If the output of the power converter (usually a SMPS) is connected to the chassis ground, no CM current flows out of the cabinet. Also, no CM currents should flow through the internal circuits. The electromagnetic interference (EMI) mains filter is easy to optimize because the high-frequency (HF) resonances are reproducible. The CM paths are mainly primary to chassis ground (through the heat sink stray capacitance) and primary to secondary (through the transformer stray capacitance).

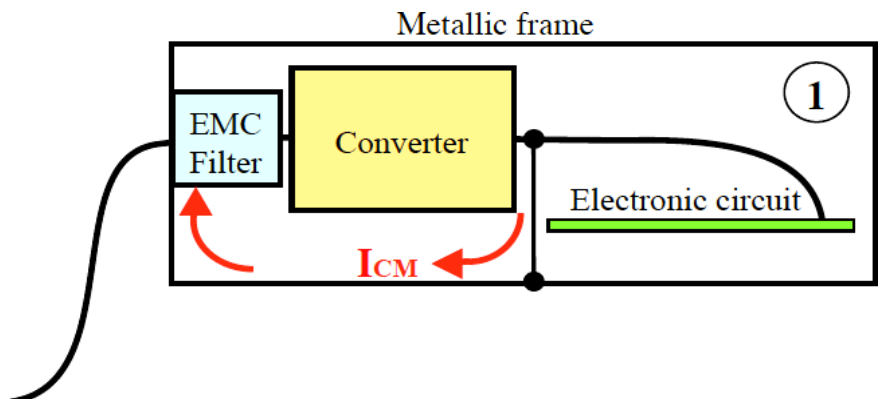

**Fig. 8:** Case 1: Best case—the converter output is grounded

### 2.6.2 Case 2: The converter output is floating, but no other cable goes out

If the output of the converter is floating, no CM current flows out of the cabinet. Nevertheless, some current may flow through the internal circuits and generate some noise. The EMI mains filter is harder to optimize because HF resonances depend on the equipment cabling (so are difficult to control, due to input-to-output variable HF impedance).

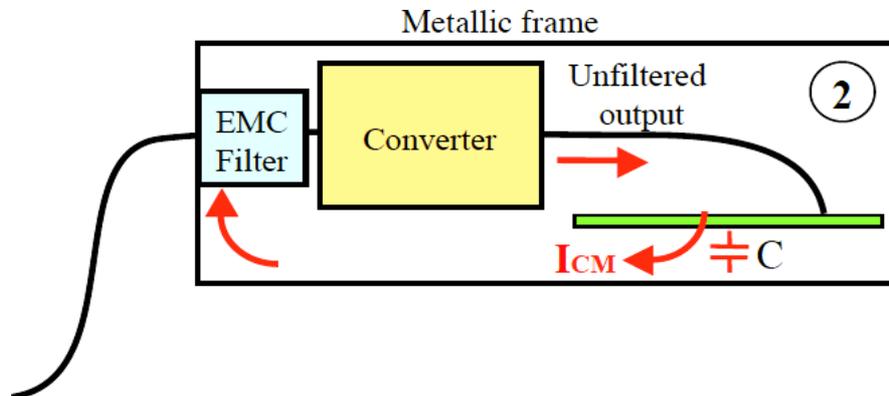

**Fig. 9:** Case 2: The converter output is floating but no other cable is routed out of the cabinet

### 2.6.3 Case 3: A noisy cable goes out of the cabinet

If a 'dirty' cable goes out of the cabinet (for instance, the cable from a variable frequency drive to its motor or to the battery of an inverter), then concerns may be severe.

First of all, the radiation of the equipment may be large and may upset other apparatus in the vicinity. Furthermore, the EMI filter is impossible to optimize because the CM currents return by any cable, even a perfectly filtered one.

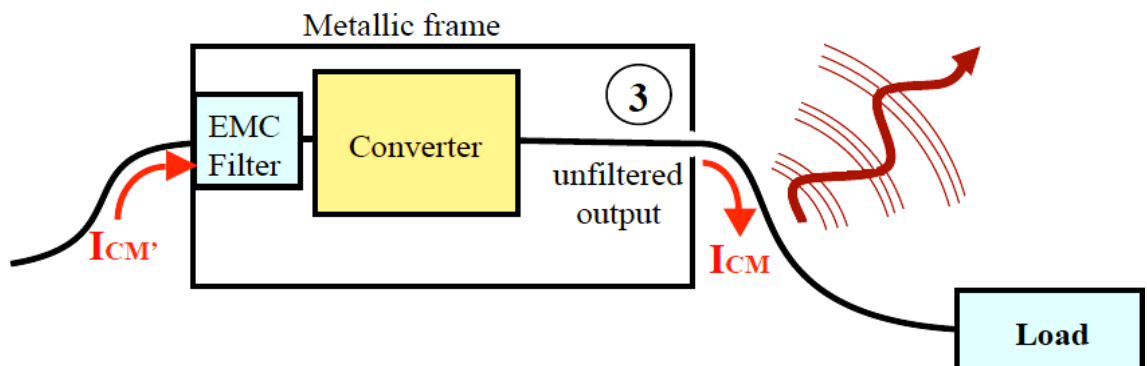

**Fig. 10:** Case 3: Worst case—a noisy cable is routed out of the cabinet

The usual solutions are:
- shield the noisy cable (with the braid connected to chassis ground at both ends),
- filter the cable (if possible, because such a filter may be a huge one).

## 3 EMC precautions in the design stage

Most of EMC concerns are easy to avoid during design. Let us consider the three most usual precautions.

## 3.1 Converter topology analysis

The easiest way to understand this is to take the example of a UPS schematic diagram.

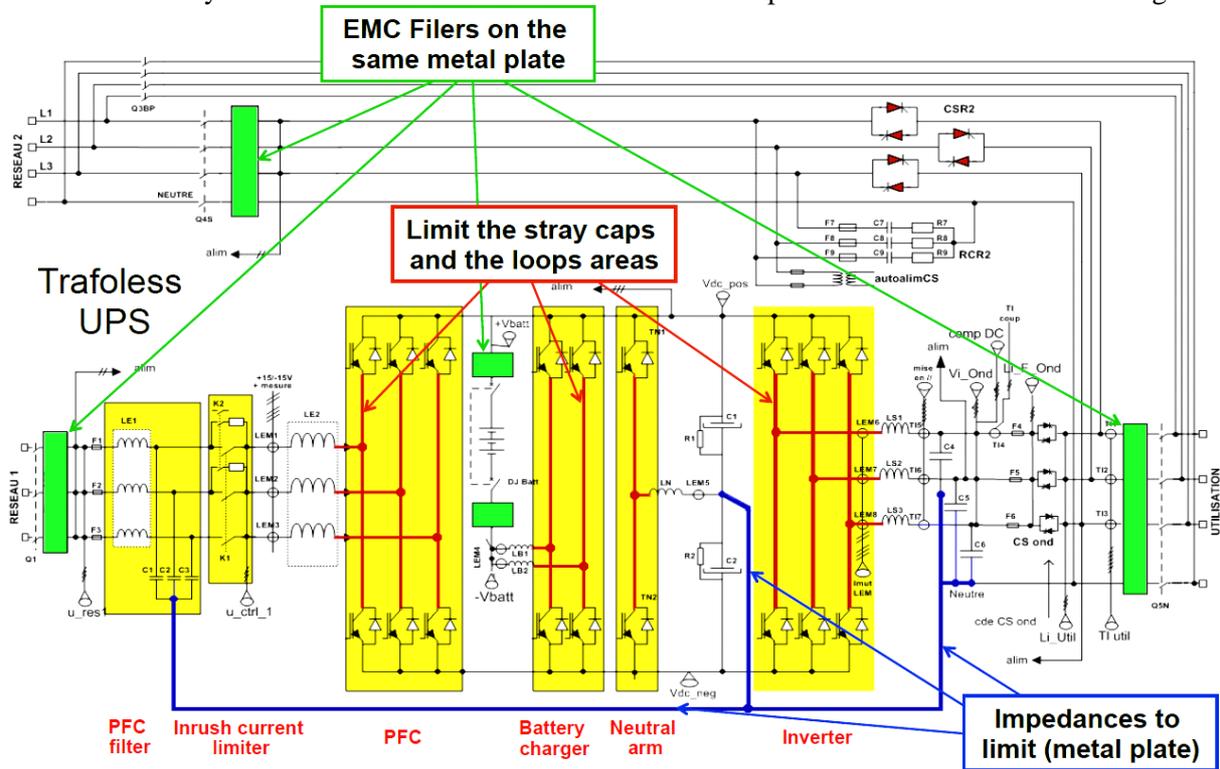

**Fig. 11:** Example of UPS schematic diagram

Figure 11 shows the schematic of a classical 'trafoless' (without an internal transformer) three-phase UPS. Three points need to be seriously controlled.

### 3.1.1 The 'hot' conductors

A 'hot' conductor is a piece of circuit submitted to high d$V$/d$t$, to high d$I$/d$t$, or both. First, we need to identify all of them (essentially the H-bridges in the upper part of Fig. 11).

The stray capacitance between high d$V$/d$t$ and chassis ground must be reduced to reduce CM current generation by:

- suppressing any localized capacitor,
- floating 'hot' heat sinks,
- increasing the distance to chassis ground,
- adding 'electrostatic' shields connected to the d.c. bus (not to ground).

The loop area of high d$I$/d$t$ must be reduced to limit H-field radiation by:

- using bus bars,
- using sandwich structures,
- reducing the size of the loops.

### 3.1.2 The 'equipotential' conductors

An 'equipotential' conductor is a conductor that must create a very low voltage drop despite a significant current in it. We must identify all of them. An example in Fig. 11 is the neutral conductor.

An equipotential conductor must be:
- wide (avoid wires, straps are preferred),
- short (place the connected components close to each other),
- of sufficient thickness—the best is a metal plate, but a short copper bar may be used.

### 3.1.3 The external cables

Any external cable must be identified and properly installed during EMC testing. Correct filtering effectiveness needs at least:
- an EMI mode filter on any external unshielded cable,
- a filter with the correct structure (for instance, without saturated chokes),
- all the filters should connect directly to the same chassis ground metal plate (to limit CM emission),
- to control the crosstalk between the 'dirty' and 'clean' sides of all those filters.

## 3.2 Cabling control

In HF, the geometry of the cabling greatly affects EMI emission. One major trap is crosstalk between the upstream and downstream of an EMI filter. It is possible to improve the effectiveness of a filter by adding capacitors, but if a star-grounding scheme is used the crosstalk between the branches of the star makes the situation worse than without added capacitors.

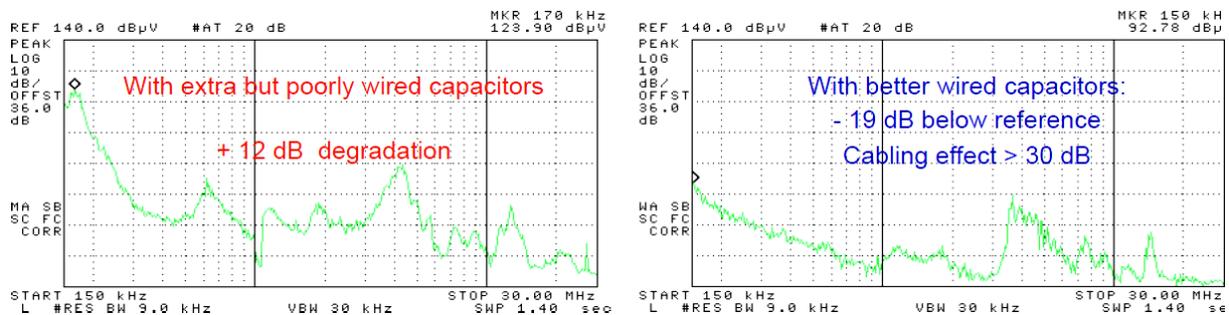

**Fig. 12:** Example of conducted EMI with poor (left) and better cabling geometry (right)

## 3.3 Risk of oscillations

Any H-bridge may oscillate if it is poorly wired or poorly controlled. Such an oscillation typically appears between 30 and 300 MHz. The effects are a loss and radiated emission increase. The risks of oscillations increase with:
- the gate trace (or cabling) length,
- the d.c. bus voltage (typically over 100 V),
- a fast d$V$/d$t$ or d$I$/d$t$ switching rate (due to a high control current),
- a fast transistor (with low parasitic capacitors),
- a low (or no) reverse blocking voltage.

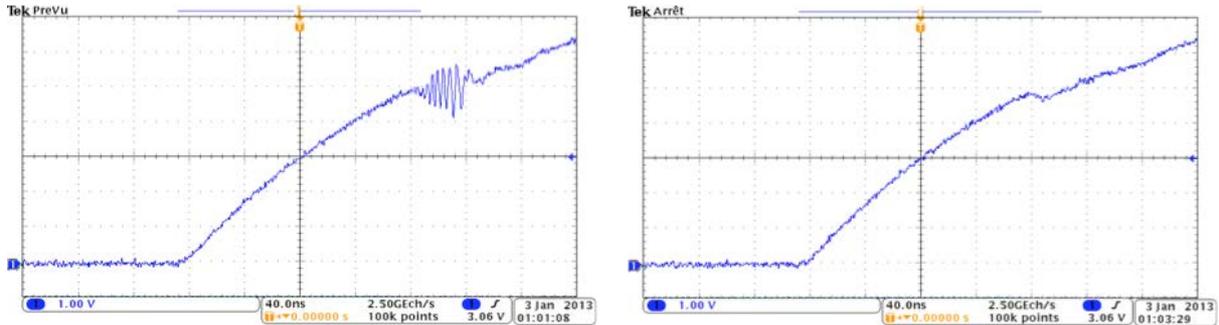
**Fig. 13:** Example of oscillation at 200 MHz (left) stopped with a ferrite bead in series with gate (right)

Solutions to an oscillation concern may be rather simple:
- the addition of a small resistor or a ferrite bead in series with the gate,
- the addition of a push–pull stage close to the gate,
- the use of a negative voltage block.

### 3.4 Immunity testing

The design of a power converter needs to control the thermal behaviour, the dynamic margins, the transient stability, and EMI emission. Those checks are usually done well. But at least two immunity tests should be done:
- a surge immunity test according to IEC 61000-4-5 standard (lightning immunity),
- an electrical fast transient in burst (EFT/B) test according to 61000-4-4 standard.

The low-frequency surge immunity may be simulated if we pay attention to non-linear effects (at least saturation of inductances and voltage clamping by MOV or Transzorb).

The HF EFT/B test cannot be correctly simulated, so a test is needed. This CM test, over each external cable, is simple, reproducible, and revealing.

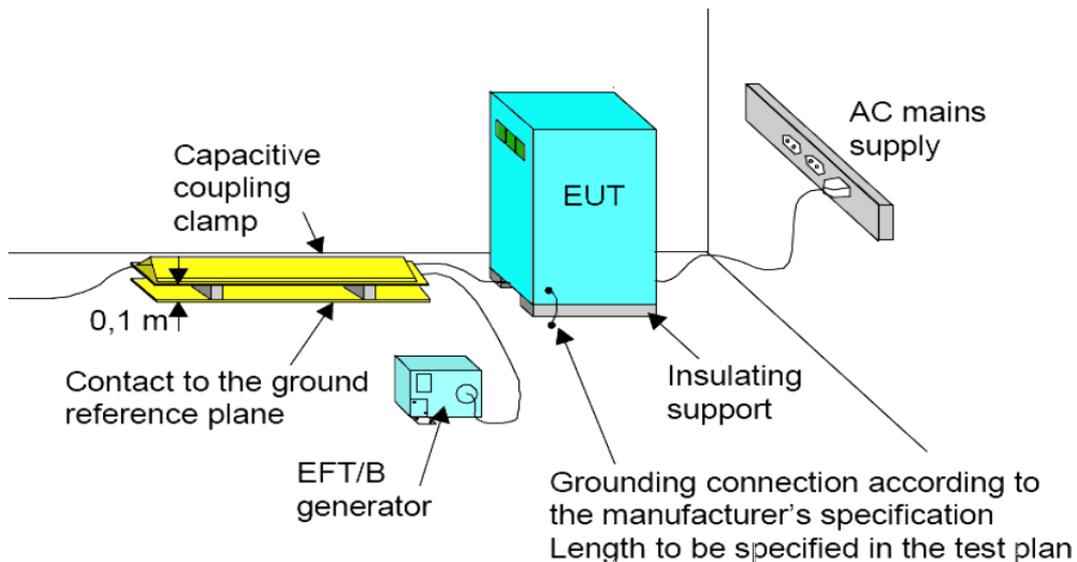
**Fig. 14:** Typical EFT/B test setup

# 4 Measuring equipment

## 4.1 Measuring equipment for development

Power converter development essentially uses time-domain measurement apparatus (data channel).

### 4.1.1 Digital oscilloscope

A digital oscilloscope is needed with, at least, the following features:

- bandwidth $\geq$ 300 MHz,
- single shot sampling frequency $\geq$ 1 Gs/s,
- fine and precise triggering threshold,
- with the possibility of screenshot.

### 4.1.2 Current probes

During development, two current probes may be used:

- active d.c. (Hall effect) current probes (if possible with a bandwidth $\geq$ 60 MHz),
- passive HF current clamp (if possible with a bandwidth $\geq$ 300 MHz).

Both can be used without any particular problems; we just have to pay attention to the risk of saturation and verify that the bandwidth is sufficient.

### 4.1.3 Voltage probes

During development, we need an active differential probe with, at least, the following features:

- primary voltage swing $\geq$ 1000 V,
- bandwidth $\geq$ 100 MHz (so a 50 $\Omega$ output impedance is needed),
- CM rejection ratio > 50 dB at 1 MHz ($\geq$60 dB recommended),
- division factor of 1/200 to 1/2000,
- noise as low as possible, but <1 V peak-to-peak brought to high-voltage side.

Unfortunately, most differential voltage probes do not meet all these requirements.

## 4.2 Measuring equipment for EMC testing

EMC testing and validation essentially uses frequency-domain data channels.

### 4.2.1 Spectrum analyser or measuring receiver

A spectrum analyser is a fast, easy-to-use, and reasonably low-cost piece of equipment. Unfortunately spectrum analysers may saturate easily and usually do not display log-log results (dB vs. log of frequency). With its pre-selector and numerous simultaneous detectors, an EMI receiver may be preferred. In both cases, the following features are required:

- input overvoltage protection (both are fragile),
- minimum frequency band 150 kHz to 1 GHz,
- MIL or CISPR normalized bandwidths.

### 4.2.2 LISN and 1500 Ω probe

Up to 16 A per phase, a Line Impedance Stabilization Network (LISN or artificial network) is required. For high currents, however, a CISPR 1500 Ω probe is required.

The implementation of a LISN requires the taking of some precautions:

- an isolation transformer to limit current leakage at 50 Hz,
- a very short ground connection to the voltage reference plate,
- voltage control of the output,
- some ventilation for high power,
- removal of the magnetic field radiated by the power converter.

A 1500 Ω voltage probe may be homemade. Its use requires the taking of some precautions:

- a short connection to chassis ground,
- the absence of interfering equipment in the environment,
- the acceptance of deviations from a measurement with an artificial network.

### 4.2.3 Current probe

For EMI measurements, we need at least two current clamps:

- A low-frequency clamp (say from 5 kHz to at least 30 MHz). This current probe should not saturated by a large 50 Hz component. Its transfer impedance may be derivative up to about 1 MHz.
- A high-frequency probe (say from 10 MHz to 300 MHz). This probe should have a flat transfer impedance. The recommended value is 5–10 Ω. Such a sensitive clamp may also be homemade.

### 4.2.4 Other near-field homemade probes

Some near-field (non-calibrated) probes are very useful for EMC analysis when developing power converters. Those E-field and H-field may be homemade. Their size will be adjusted as required.

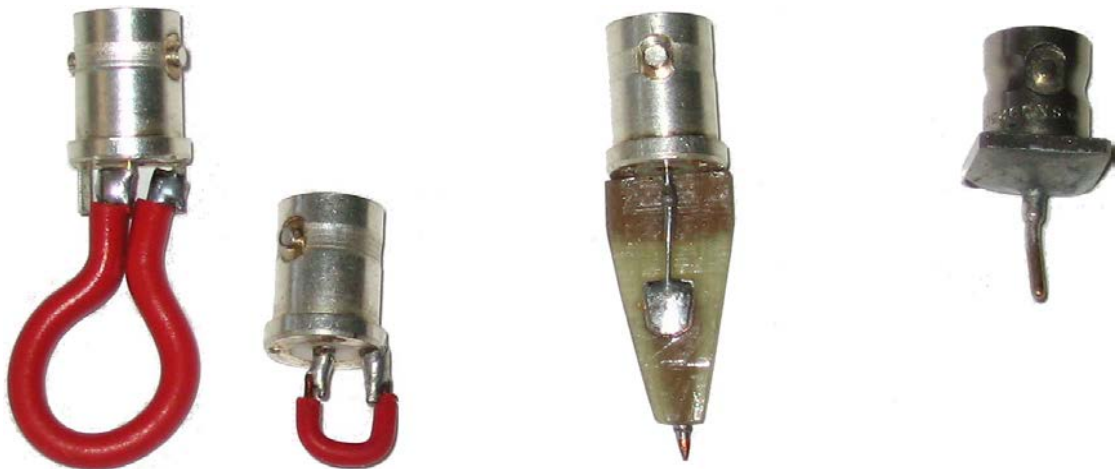

**Fig. 15:** Some examples of homemade $\Delta B/\Delta t$, $\Delta V/\Delta t$, and $Z_t$ measurement probes

# 5    Conclusions

Power converters may generate a large amount of electromagnetic noise. They must be filtered and, possibly, shielded. Some features can be correctly simulated, for instance low-frequency stability or transient behaviour. Unfortunately, some EMC features can only be measured at the end of the design. For instance, conducted or radiated behaviour over 1 MHz for a large converter (10 MHz for a small one) is quite difficult to simulate correctly.

The EMC of a power converter is quite easy to control if we correctly:

- foresee phenomena during an early part of the design stage,
- plan the time to analyse and understand the measurements after development,
- plan the time to fix the troubles and to optimize the EMC solutions (filters and shields).

Recall that most large converters (UPS or motor drives) meet European standards but not the requirements of the EMC directive. That is the reason why the CISPR standard should be reviewed:

- to suppress the 'Wild West effect' (from d.c. to 150 kHz) in both emission and immunity,
- to reduce the permitted conducted emission level for equipment over 100 A per phase.